# Superconducting Fluctuations and Anomalous Phonon Renormalization much above superconducting transition temperature in $Ca_4Al_2O_{5.7}Fe_2As_2$

Pradeep Kumar[1], A. Bera[1] D. V. S. Muthu[1], P. M. Shirage[2], A. Iyo[2] and A. K. Sood[1,*]

[1]Department of Physics, Indian Institute of Science, Bangalore -560012, India

[2]National Institute of Advanced Industrial Science and Technology, Tsukuba, Ibaraki 305-8568, Japan

**ABSTRACT**

Raman studies on $Ca_4Al_2O_{5.7}Fe_2As_2$ superconductor in the temperature range of 5 K to 300 K, covering the superconducting transition temperature $T_c \sim 28.3$ K, reveal that the Raman mode at $\sim 230$ cm$^{-1}$ shows a sharp jump in frequency by $\sim 2$ % and linewidth increases by $\sim 175$ % at $T_o \sim 60$ K. Below $T_o$, anomalous softening of the mode frequency and a large decrease by $\sim 10$ cm$^{-1}$ in the linewidth is observed. These precursor effects at $T_0$ ( $\sim 2T_c$ ) are attributed to significant superconducting fluctuations, possibly enhanced due to reduced dimensionality arising from weaked coupling between the well separated ($\sim 15$ Å) *Fe-As* layers in the unit cell. A large blue-shift of the mode frequency between 300 K to 60 K ( $\sim 7$ % ) indicates strong spin-phonon coupling in this superconductor.

*Corresponding author: email: asood@physics.iisc.ernet.in, Ph: +91-80-22932964

# 1. INTRODUCTION

The discovery of high temperature superconductivity in iron-pnictide materials[1] has generated enormous interest, both experimentally and theoretically. Among the iron-pnictides superconductors discovered recently are systems having identical FePn (Pn = As, P) layers but with various block layers.[1-3] The families of materials which have block layers of the perovskites structure allow a controlled tailoring of the distance between the neighboring FePn layers, such as in $Ba_4Sc_3O_{7.5}Fe_2As_2$,[4] $Ca_{n+m}(M,Ti)_nFe_2As_2$ (M = Sc, Mg, Al; n = 2,3,4,5 and m = 1, 2),[5-6] $Ca_3Al_2O_{5-x}Fe_2Pn_2$ (Pn = As, P)[7] and $Ca_4Al_2O_{6-x}Fe_2Pn_2$ (Pn = As, P),[8] with transition temperature up to 47 K. Of these, $Ca_3Al_2O_{5-x}Fe_2Pn_2$ and $Ca_4Al_2O_{6-x}Fe_2Pn_2$ (Pn = As, P) have the smallest $a$-lattice parameters and the largest iron interlayer distance ($d_{Fe}$). Infact emergence of superconductivity in these compounds has been ascribed to the smallness of the tetragonal $a$-axis lattice constant.[7-8] It has been further shown that increasing $d_{Fe}$ is correlated with the enhanced $T_c$, possibly due to a reduced coupling between different Fe-As layers.[9] Since Fe-As planes are believed to be the key players in the occurrence of superconductivity in iron pnictides, a large $d_{Fe}$ would cause the dimensionality to decreases from three (*3D*) to two (*2D*) dimensions. As a consequence, the superconducting fluctuations above $T_c$ should be enhanced-an issue which has been earlier addressed in high temperature cuprates superconductors, but not in iron based superconductors. In this work, we demonstrate the onset of superconducting fluctuations at ~ 60 K, i.e at ~ *2$T_c$,* using Raman phonon renormalization finger prints.

In superconductors, the opening of a superconducting gap renormalizes the electronic states near the Fermi surface, which in turn can change the phonon self-energies. Correspondingly, Raman and infrared spectra do reveal anomalies in vibrational modes below $T_c$ attributed to an opening of a superconducting gap. These superconductivity-induced self-energy effects have been used to estimate the magnitude and symmetry of the superconducting order parameter.[10-14] On the other hand, precursor effects of superconducting fluctuations above $T_c$ on phonons can provide information on a relatively local scale in contrast to transport studies which probe average bulk properties. We here show that phonon anomalies seen at a temperature of ~ $2T_c$ imply the existence of superconducting clusters or droplets much above the critical temperature $T_c$, consistent with the Tisza-London model of a superconductor.[15] The precursor effects above $T_c$ arise from strong fluctuations of the phase of the complex superconducting order parameter thus suggesting the existence of phase-incoherent cooper pairs above $T_c$. Within the model of phase-incoherent cooper pairs originating from the classical fluctuations of the phase of the superconducting order parameter, pairing of the superconducting quasiparticles is expected to start much above $T_c$ ; however, global phase coherence is achieved only below $T_c$. The superconducting transition in classic superconductors is much sharper because the coherence length is much larger than the interatomic distance. However, if coherence length is comparable to atomic dimensions, then it will lead to much more prominent fluctuation effects. Iron pnictides have small coherence length similar to cuprates superconductors,[16] thus setting the stage for superconducting fluctuations.

As mentioned earlier, a few temperature dependent Raman studies have been reported for "1111", [10, 17] "122" [11, 18-19] and "11" [20-21] systems to study about the effect of superconducting transition on the phonons below $T_c$. However, no precursor effects have been seen in these studies. In addition, to the best of our knowledge, there are no reports so far of Raman study of the new compound $Ca_4Al_2O_{6-y}Fe_2As_2$ (referred as Al-42622). In this paper we report Raman study of $Ca_4Al_2O_{5.7}Fe_2As_2$ with $T_c$ ~ 28.3 K [8] in the temperature range 5 K to 300 K covering the spectral range of 120 cm$^{-1}$ to 800 cm$^{-1}$. We present two significant results: (i) the Raman mode at ~ 230 cm$^{-1}$ show a jump in frequency by ~ 5 cm$^{-1}$ ( ~ 2 % ) and the corresponding linewidth increases from ~ 16 cm$^{-1}$ to ~ 44 cm$^{-1}$ ( ~ 175 % ) at temperature $T_o$ ~ 60 K. The mode frequency shows anomalous softening by ~ 5 cm$^{-1}$ below $T_o$, accompanied by a large decrease of linewidth by ~ 10 cm$^{-1}$ from 60 K to 5 K. These results on phonon renormalization with an onset at ~ $2T_c$ are attributed to superconducting fluctuation - induced effects on the phonon self-energy.

(ii) A large increase of the phonon frequency (~ 7 %) between 300 K and 60 K suggests a strong spin-phonon coupling in $Ca_4Al_2O_{5.7}Fe_2As_2$, similar to other iron based superconductors.[11, 18, 20]

## 2. EXPERIMENTAL DETAILS AND RESULTS

Polycrystalline samples with nominal composition of $Ca_4Al_2O_{5.7}Fe_2As_2$, and with a superconducting transition temperature of $T_c$ ~ 28.3 K were prepared using high pressure synthesis. The samples were characterized as per details described in Ref. 8. Unpolarised micro-Raman measurements were performed similar to our earlier Raman studies on iron-pnictides.[10-11] $Ca_4Al_2O_{5.7}Fe_2As_2$ has a layered structure belonging to the tetragonal *P4/nmm* space group. The inset in Fig.1 exhibits the Raman spectrum at 5 K revealing only one mode. Raman spectra shown in Fig.1 at a few temperatures are fitted with a Lorentzian function indicated by solid lines. Figure 2 shows the peak frequency $\omega$ (bottom panel) and full width at half maximum (FWHM) (top panel) as a function of temperature. The following observation can be made: (i) The frequency of the mode shows anomalous hardening ( ~ 7 % ) between 300 K and $T_o$ ( ~ 60 K ), (ii) the frequency jumps abruptly by ~ 5 cm$^{-1}$ at $T_o$. (iii) The mode frequency displays a softening below $T_o$ (iv) The linewidth decreases slightly ( ~ 2.5 cm$^{-1}$) between 300 K to 60 K, as expected due to anharmonic interactions, and shows a very large anomalous increase ( ~ 175 % ) at $T_o$. (v) Below $T_o$, the linewidth decreases significantly from ~ 46 cm$^{-1}$ to 35 cm$^{-1}$ in a narrow temperature range from 60 K to 5 K.

## 3. DISCUSSION

The anomalous hardening of the mode from room temperature to 60 K ( ~ 7 % ) may be attributed to strong spin-phonon coupling, similar to other studies for similar iron-based systems. The coupling between phonons and the spin degrees of freedom can arise either due to modulation of exchange integral by phonon amplitude and/or by involving change in the Fermi surface by spin-waves, provided the phonon couples to that part of the Fermi surface.[22-23] We note that earlier [11, 18, 20] Raman studies on other iron-based superconductors have also indicated strong spin-phonon coupling, supported by theoretical calculations.[24]

The changes in the frequency and linewidth of phonons across the transition temperature $T_c$ are due to changes in the phonon self energy, $\Delta\Sigma = \Delta\omega + i\Delta\Gamma$ induced by the superconducting transition.[25] The real and imaginary parts of the self energy renormalize the frequency and the life-time of the phonon, respectively. Within the framework of strong coupling Eliasberg theory, Zeyer et al [25] showed that a change in the phonon self-energy across the transition temperature is linked with the interaction of the phonons with the superconducting quasiparticles excitations. Qualitatively, phonons with frequency below the superconducting gap ($2\Delta$) can soften whereas the phonons with frequency above the gap can harden. The renormalization of a given phonon mode is strongest when the phonon energy is close to the gap. However, Fig. 2 clearly shows that the onset of renormalization of the Raman phonon is at a temperature $T_o$ much higher than $T_c$. We note that phonon anomalies have been reported for cuprate superconductors at temperature as high as ~ $2T_c$ attributed to the superconducting order-parameter fluctuations.[26] The fact that precursor effects seen in this work have not been observed in other iron-based superconductors is possibly related to the reduced dimensionality of the Cooper pair wavefunction due to large distance $d_{Fe}$ between the two Fe-As layers in the unit cell. The large anomalous softening of the frequency at $T_o$ and significant broadening of the observed mode suggests that the phonon frequency is very close to the superconducting gap (i.e $2\Delta$ ~ 230 cm$^{-1}$) in superconducting clusters existing above $T_c$. Near $T_o$, opening of superconducting gap provides additional decay channel due to the breaking of the cooper pairs leading to an increase of the phonon linewidth. On further decreasing the temperature the phonon is inside the gap, which results in sharpening of the linewidth as decay channels are being removed.[13-14] The anomalous softening below $T_o$ and a large jump in the broadening of the mode at $T_o$ are evidence of a strong coupling between the Raman phonon and the superconducting quasiparticle excitations.

The importance of incoherent phase fluctuations may be understood by using experimentally determined quantities to evaluate the characteristic temperature $T_\theta$ which is linked to the stiffness of the phase of the superconducting order parameter. For $T_c \sim T_\theta$, the incoherent phase fluctuations are prominent even for $T > T_c$ but no global phase coherence till $T_c$ to give bulk superconductivity. The characteristic energy scale for phase fluctuations of the superconducting order parameter is the zero temperature phase stiffness given as $V = (\hbar c)^2 a/(16\pi e^2 \lambda_o^2)$, where 'a' is the length scale, defined as the average spacing between superconducting layers, and $\lambda_o$ is the penetration depth. The characteristic temperature ($T_\theta$) is related to the phase stiffness as $T_\theta = AV$ ($A \sim 1$).[27] Taking $\lambda_o \sim 200$ nm,[16,28] we estimate that $T_\theta \sim T_c$, similar to that in cuprates superconductors.[27] This suggests that incoherent phase fluctuations with the non-zero value of the superconducting gap are possible above $T_c$.

## 4. CONCLUSIONS

In conclusion, we have provided evidence for strong coupling of the observed phonon mode with the superconducting order parameter fluctuations setting in at almost ~ $2T_c$. Strong precursor effects due to existence of dynamic superconducting clusters at temperatures as high as $2T_c$ are attributed to the reduced overlap of the electronic wavefunction of the Fe-As layers in this compound due to large distance between the pnictides layers. It will be interesting to look for these strong precursor effects in careful transport measurements and Andreev reflection in scanning tunneling microscopy.

**Acknowledgments**

PK, AB acknowledges CSIR, India, for a research fellowship and AKS acknowledge DST India for financial support. We thank Prof. S. Dattagupta and Prof. Pratap Raychaudhari for very useful discussion. PMS and AI acknowledge financial support by Grant-in-Aid for specially promoted Research (20001004) from the Ministry of Education, Culture, Sports, Science and Technology (MEXT) and thanks Dr. H. Eisaki and Y. Tanaka for fruitful discussions.

**FIGURE CAPTION:**

FIGURE.1 (Color online) Raman spectra of $Ca_4Al_2O_{5.7}Fe_2As_2$ at a few temperatures. Solid lines are the fit to the Lorentzian function. Inset shows the spectrum at 5 K in a large spectral window.

FIGURE.2. (Color online) Temperature dependence of the phonon mode. Solid lines from 300 K to $T_o$ are linear fits and dash lines are guide to the eye.

FIGURE1:

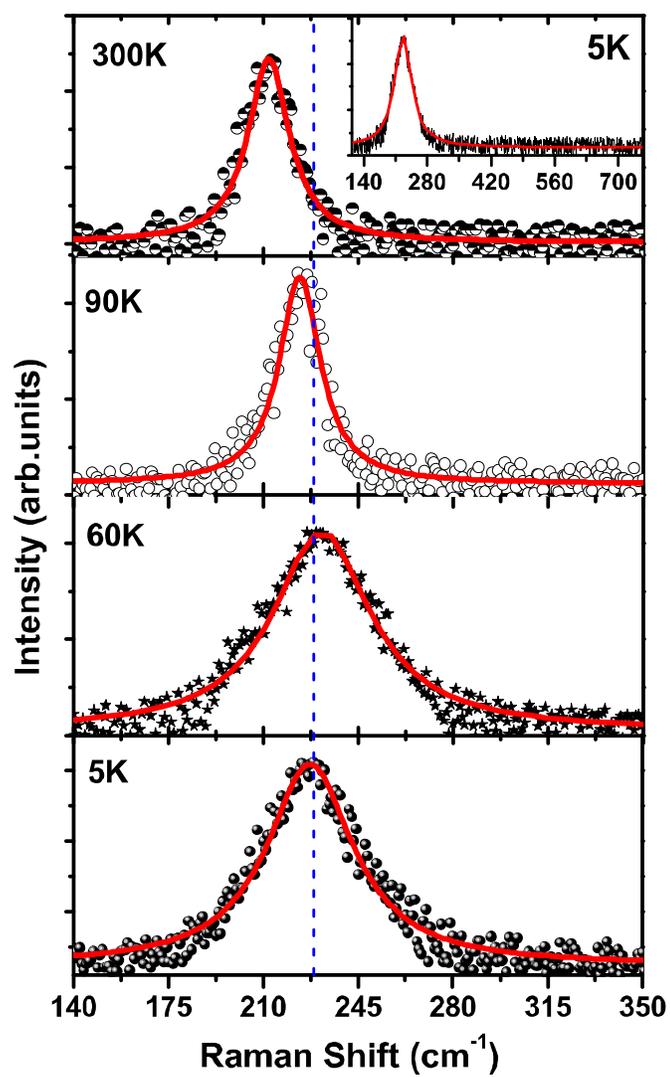

FIGURE 2:

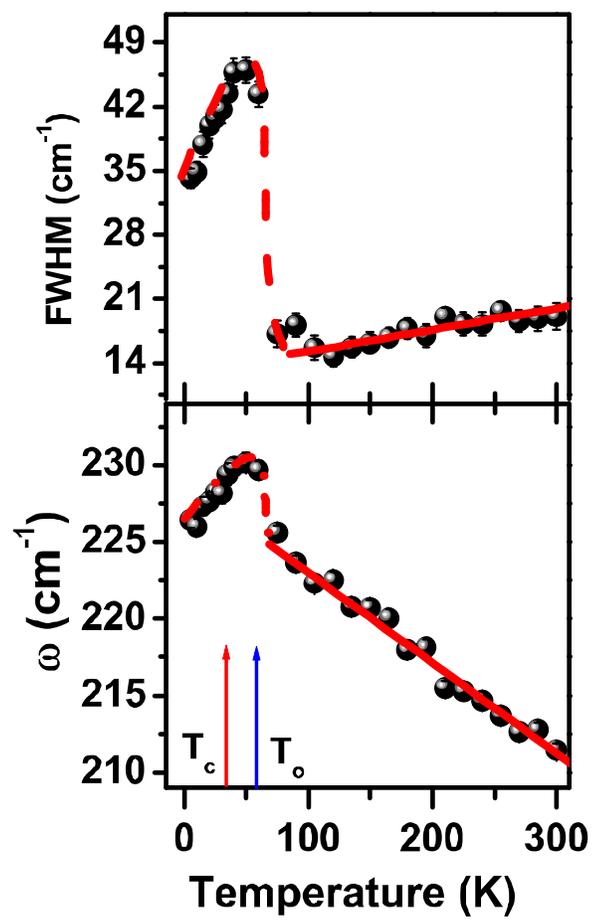